\documentclass[twocolumn]{aastex631}

\shorttitle{A Cacophony of Echoes}
\shortauthors{Serafin Nadeau et al.}

\graphicspath{{./}{figures/}}

\begin{document}

\title{A Cacophony of Echoes from daily monitoring of the Crab Pulsar at Jodrell Bank}

\correspondingauthor{Thierry Serafin Nadeau}
\email{serafinnadeau@astro.utoronto.ca}

\author[0009-0005-6505-3773]{Thierry Serafin Nadeau}
\affil{David A. Dunlap Department of Astronomy and Astrophysics,
50 St. George Street,
Toronto, ON M5S 3H4, Canada}
\affil{Canadian Institute for Theoretical Astrophysics,
60 St. George Street,
Toronto, ON M5S 3H8, Canada}

\author[0000-0002-5830-8505]{Marten H. van Kerkwijk}
\affil{David A. Dunlap Department of Astronomy and Astrophysics,
50 St. George Street,
Toronto, ON M5S 3H4, Canada}

\author[0000-0002-1429-9010]{Cees G. Bassa}
\affil{ASTRON, Netherlands Institute for Radio Astronomy,
Oude Hoogeveensedijk 4, 7991 PD, Dwingeloo, The Netherlands}

\author[0000-0001-9242-7041]{Ben W. Stappers}
\affil{Jodrell Bank Centre for Astrophysics, Department of Physics and Astronomy, University of Manchester, Manchester M13 9PL, UK}

\author[0000-0001-6798-5682]{Mitchell B. Mickaliger}
\affil{Jodrell Bank Centre for Astrophysics, Department of Physics and Astronomy, University of Manchester, Manchester M13 9PL, UK}

\author[0000-0002-4799-1281]{Andrew G. Lyne}
\affil{Jodrell Bank Centre for Astrophysics, Department of Physics and Astronomy, University of Manchester, Manchester M13 9PL, UK}

\begin{abstract}

Using archival data from the 42 foot telescope at the Jodrell Bank Observatory, we produce daily stacks of aligned giant pulses for the Crab pulsar, to study changes to the daily profiles between April 2012 to December 2016. From these, we identify echoes, where intervening material away from the line of sight causes pulsed emission to be redirected towards the observer, with delay corresponding to the increased distance of travel, resulting in additional profile components.
These observations show that such echoes may be far more common than implied by the previous rate of detections.
All the observed echoes are consistent with approaching zero-delay at their closest approach to the normal giant pulse emission.
This indicates that the structures responsible for producing these events must be highly anisotropic, with typical lengths greater than $\sim 4\textrm{AU}$, typical widths on the sky of $\sim 0.1 \textrm{AU}$ and typical depths of $\sim 5\textrm{AU}$, given the previously observed electron densities of the nebular filaments, on the order of 1000 cm$^{-3}$.
This suggests that these inhomogeneities are likely to be offshoot substructure from the larger nebular filaments of the Crab nebula.

\end{abstract}


\section{Introduction}\label{sec:intro}

The Crab pulsar is among the 11 pulsars known to exhibit brief flashes of extraordinarily bright radio emission known as ``giant pulses'' \citep{kuzmin07}.

These giant pulses do not occur every rotation and are usually confined a narrow phase window, often trailing a location where more normal pulsed emission is seen (e.g., the precursor component in the Crab).
For the Crab, these are  the Main Pulse (MP) and Interpulse (IP), which dominate its overall pulse profile \citep{hankins07}.
The Crab's giant pulses are short in duration, of the order of a few micro-seconds or less, and have been found to consist of nanosecond-duration ``nano shots'' \citep{hankins03}.

As pulsar radio emission travels to us from its source, it is affected by its interaction with the ionized plasma of the interstellar medium, or ISM.
Hence, we observe the convolution of the intrinsic pulsar signal with the Impulse Response Function (IRF) of the intervening material, both via dispersion, which can be corrected for, and by multipath scattering, which cannot (see \cite{mckee18} for a long-term perspective of scattering of the Crab pulsar).
For giant pulses, given their narrow width, the observed signal is dominated by the IRF, and thus they serve as excellent probes of interstellar and nebular structures along the line of sight.

As it turns out, on top of being one of the few pulsars with giant pulses, the Crab is also one of the few pulsars known to sometimes exhibit echoes in its pulse profile: copies of the profile which are delayed in time relative to the regular pulse profile.
% Typically during these events, additional components in the Crab's emission appear, with a time delay from the giant pulses.
For the Crab, a particularly strong event happened in 1997, which lasted on the order of $\sim\!100{\rm\;d}$ \citep{backer00, lyne01}.
This event made clear that the echoes are caused by the line of sight passing near structures in the nebula, with the amplitude increasing and delay decreasing as the line of sight and the structures approach one another.
However, the mechanism underlying the echoes remains uncertain.
Originally, \cite{grahamsmith00} and \cite{backer00} proposed that it was specular reflections off the surface of an intervening ionized bubble in the Crab nebula.
Later, \cite{grahamsmith11} suggested that refraction within these clouds was a more likely mechanism.

After the identification of the 1997 event with an echo, similar echoes have been identified retroactively, in 1974 \citep{lyne75}, 1992 and 1994 \citep{lyne01}, as well as in observations since (e.g., \citealt{driessen19}).
These further echoes are less strong and seem to persist for shorter periods of time compared to the 1997 event, with timescales of a few days days to a few weeks \citep{crossley07, driessen19}.
The delay seen in the echoes also varies significantly, with \cite{crossley07} finding delays spanning 40 to $100{\rm\,\mu s}$, while \cite{driessen19} and \cite{grahamsmith11} find delays of 1.5 and $5{\rm\,ms}$, respectively.

As a result of their typically brief duration and relatively irregular occurrence, echoes are unlikely to be consistently observed in their entirety with random observations of the Crab pulsar.
Instead, regular observations over longer periods of time are required.

Here, we use Crab observations with the 42 foot telescope at the Jodrell Bank Observatory for this purpose.
We focus on the giant pulses, which, given their short intrinsic duration and high intensity, serve as very good probes of the Impulse Response Function of the structures causing the echoes, as long as one ensures that they are carefully aligned before averaging, so that one avoids the smearing induced by the fact that giant pulses occur randomly within $\sim\!1\%$ of pulse phase.

We describe the observations and the selection of giant pulses in Sect.~\ref{sec:selection}.
In Sect.~\ref{sec:stacking}, we describe how they are aligned and averaged into daily stacks, to gain sufficient signal-to-noise to detect also fainter echoes (with the daily timescale still well below the timescale on which echoes evolve).
In Sect.~\ref{sec:echoes}, we show the results and analyse identified echoes.
We discuss the implied locations and physical properties of the scattering regions in Sect.~\ref{sec:structures}, and the ramifications of the work in Sect.~\ref{sec:rams}.

\section{Observations and Giant Pulse Selection} \label{sec:observations}\label{sec:selection}

\begin{figure*}
  \centering
  \includegraphics[width=0.46\textwidth]{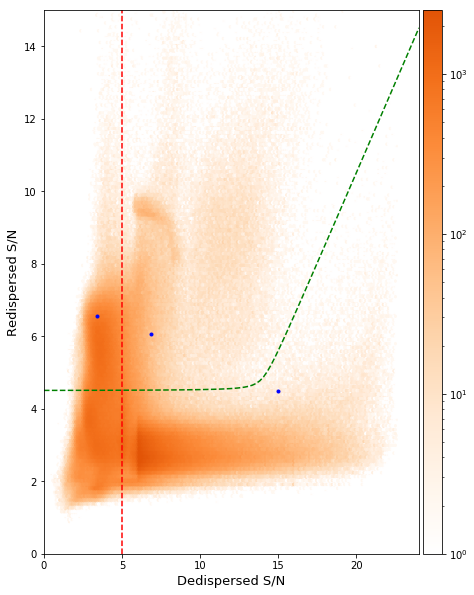}
  \includegraphics[width=0.49\textwidth]{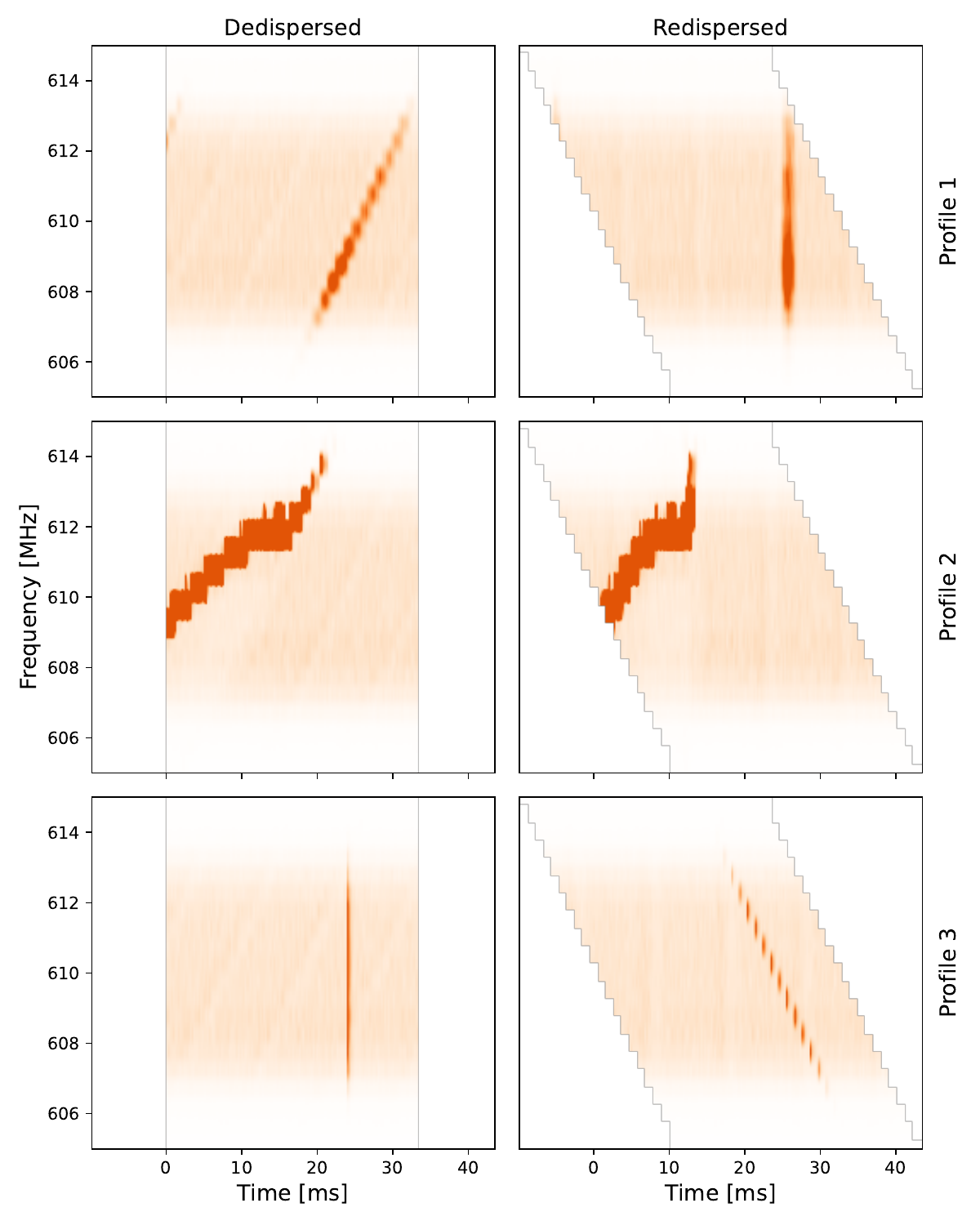}
  \caption{
    Identifying genuine giant pulses.
    {\em Left:\/} Peak signal-to-noise ratio in the redispersed versus the dedispersed time stream for all triggers in our data set.
    The red and green dashed lines indicate the left and upper boundaries of our selection criterion, removing pulses that are too faint or that are likely due to broadband RFI -- which generally has higher signal-to-noise in the redispersed timestream.
    {\em Middle and Right Panels:\/} Dedispersed and redispersed timestreams for the three triggers marked with blue dots.
    One sees that the first two are better aligned in the redispersed timestream, suggesting they are due to RFI.
    In contrast, the third trigger is clearly aligned across frequency, and thus likely is a giant pulse.
    \label{fig:sn}\label{fig:S/Nselection}\label{fig:rfi}
  }
\end{figure*}

Since 1984, the Crab pulsar has been monitored with the 42 foot telescope at the Jodrell Bank Observatory nearly any time it is visible, mostly to track the evolution of its spin frequency \citep{lyne15}.
This work is based on 1721 days of observations, from 2012 April 16 to 2017 January 1 (MJD 56033--57754).

The observations used were obtained with the \texttt{COBRA2} backend, installed in 2012, which records Nyquist sampled, dual circular polarization, complex voltages for 10\,MHz of bandwidth centered at 610\,MHz, with a strong dropoff in signal outside the central 5\,MHz due to an RF filter. The resulting complex voltages were coherently dedispersed at the nominal dispersion measure (DM) of the Crab pulsar of $\sim\!56.8{\rm\,pc/cm^3}$.  The data were channelized into 0.5\,MHz channels using \texttt{dspsr} \citep{vanstraten11} and folded using the regularly updated Jodrell Bank Crab ephemerides \citep{lyne93}. Pulse profiles for individual rotations were processed by masking the outer 4 frequency channels and averaging each profile in frequency and polarization to 512 pulse phase bins. The standard \texttt{psrchive} \citep{hotan04} metrics were used to define the off-pulse region, and its mean and standard deviation used were to select those pulse rotations where the maximum flux value of the entire profile exceeded the noise by $6\sigma$. For those profiles, pulse profiles with full frequency and polarization information were stored at 8192 pulse phase bins, where each bin contains 2 or 3 channelized time samples.

Since the selection is based on a simple signal-to-noise (SN) threshold, the archived profiles contain many false positives due to radio frequency interference (RFI).
One common type of RFI is narrow-band, being more or less continuously present in the top two frequency channels (614--615\,MHz).
Since this RFI dominates those channels causing both false positives and substantial extra noise in genuine pulses, we mask these channels in our analysis.

A second common form of RFI is impulsive and broadband (for an example, see Figure \ref{fig:rfi}).
In the dedispersed time stream, it results in profiles that last roughly 0.01s, sweeping up in frequency, i.e, with the time delay expected from dedispersion over the 5-MHz frequency band for the pulsar (DM) of $\sim\!56.8{\rm\,pc/cm^3}$.

Since genuine giant pulses are narrow in the dedispersed time stream, we can distinguish those from the impulsive RFI by comparing the SN as measured in the dedispersed time stream with the SN inferred after redispersing the data to its original state: genuine giant pulses and RFI will have higher SN in the dedispersed and redispersed time streams, respectively.
In Figure \ref{fig:sn}, we show the SN with and without redispersion for all triggers, along with a few examples.
The real giant pulses are confined to a band with a range in dedispersed peak SN but low redispersed SN.

For our selection of giant pulses, we start by removing all low-SN triggers, with a dedispersed peak SN of less than $5\sigma$ (thus removing pulses triggered or greatly affected by the narrow-band RFI).
Next, we exclude all triggers above the green cutoff shown in Figure \ref{fig:sn}, which is flat (redispersed S/N of $4.5\sigma$) before transitioning to a slope of 1 for the brightest pulses (where the numbers were chosen empirically to exclude as many false triggers as possible while being sure to include pulses that visual inspection showed were clearly genuine).
In the end, these cuts removed 2256672 of 3038254 triggers, or about 74.3\%.

\begin{figure}
  \centering
  \includegraphics[width=0.5\textwidth]{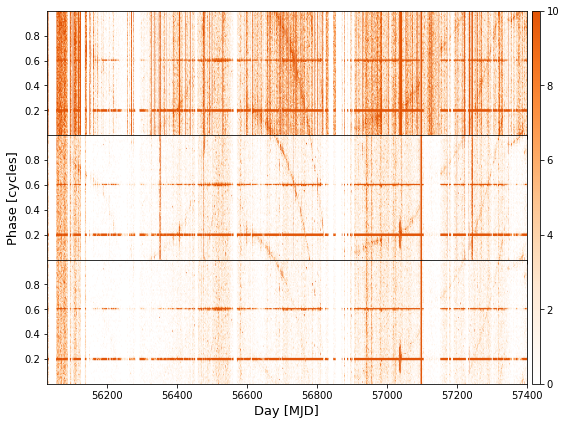}
  \caption{
    Phase distribution of all triggers ({\em top}), after removing low signal-to-noise ones ({\em middle}), and after clipping against broadband RFI ({\em bottom}), using constraints shown in Figure \ref{fig:S/Nselection}.
    One sees that a large majority of archival triggers profiles are not genuine giant pulses.
    \label{fig:phaseselection}\label{fig:phase}
  }
\end{figure}

Even with this initial SN based selection, the distribution over pulse phase shows that there are still many profiles incorrectly identified as as Crab giant pulses (see Figure \ref{fig:phase}).
In order to reduce the false-positive rate further, we ignore all triggers that are more than 1.5\% in phase away from the Main pulse and Interpulse phase windows.
For this purpose, we first convert times to pulsar phases using \verb|tempo2| \citep{hobbs06} along with the Jodrell Bank Crab ephemerides \citep{lyne93} for each sidereal day.
We find that in the resulting phase distributions, the main and interpulse wander slowly with time as the ephemeris loses its predictive power, and shows jumps when the timing ephemeris is updated.
For our purposes, though, this is not very important: for almost all days, we can use the distribution of triggers in phase to reliably determine the offsets of the Main pulse and Interpulse windows, and only a few real pulses end up being removed by our phase gating.
Lastly, during phase gating, we mask out times in which there is an excess of triggers across all phases, implying a large rate of false positives also in the pulse gates.
After the phase selection, we are left with a total of 733663 triggers.

We can estimate the remaining false-positive rate from the number of profiles which are out of phase with the giant pulses after our SN selection but not at bad times (437954 triggers). Given that these cover 94\% in phase, we infer that in each of the Main and Interpulse phase windows there will be about 14000 false positives, or a false-positive rate of 3.8\% (2.2\% and 16\% of Main Pulse and Interpulse, respectively).
This of course varies from day to day with the 1$\sigma$ range around the median being $3.1^{+10}_{-0.7}\%$.

\section{Pulse Profiles}\label{sec:stacking}

\begin{figure*}
  \centering
  \includegraphics[width=0.7\textwidth]{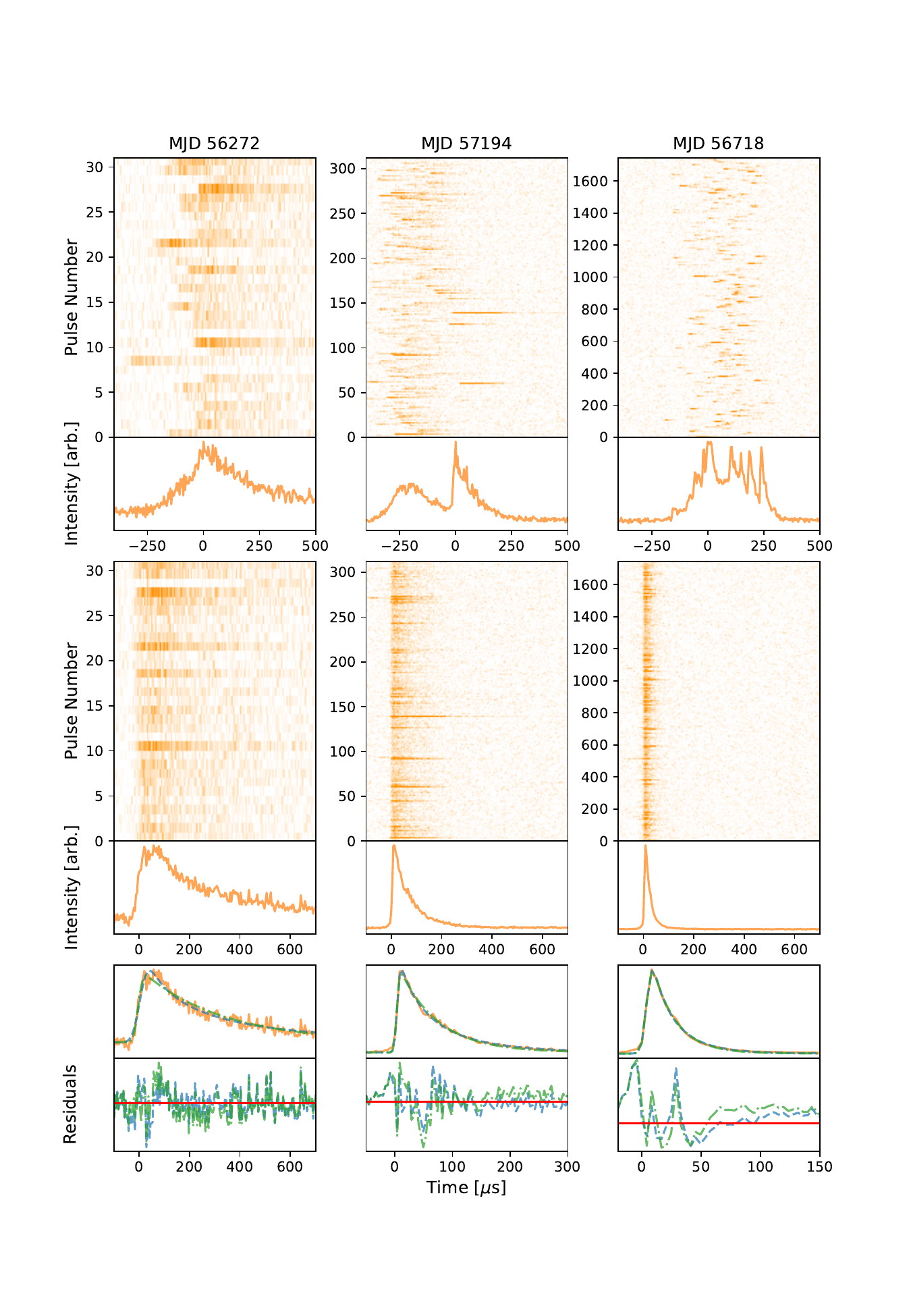}
  \caption{
    Pulse stacks, before ({\em top}) and after ({\em middle}) alignment.
    Stacks are shown for three days, with low, moderate and high number of giant pulse detections, with the resulting weighted profiles below.
    For all days, correcting for jitter in the exact phase at which a pulse occurs leads to clear improvements, with the most drastic effect seen in days with low scattering, when the phase jitter is large compared to the scattering time.
    Fits to the aligned and averaged stack profiles are also shown ({\em bottom}) with the residuals from the fits below, for both the exponential tail (green dot-dashed line) and the modified exponential tail (blue dashed line) functions.
    \label{fig:alignment}\label{fig:align}
  }
\end{figure*}

For different days, the removal of false triggers leaves us with greatly varying numbers of pulses, ranging from tens to thousands.
Generally, fewer pulses are detected on days that the pulses have very long scattering tails as their peak fluxes thus less frequently reach the trigger threshold. On days when the Sun is close to the Crab on the sky, the number of detections is also reduced due to the decreased sensitivity of the telescope.
More accurate pulse profiles are obtained by averaging pulses, but before doing that, we must take into account that pulses do not arrive at exactly the same phase, but rather randomly occur in their phase windows, i.e., up to a few hundreds of $\mu s$ apart (Figure \ref{fig:align}), a phenomenon intrinsic to the emission known as jitter.

In order to determine the individual pulse phases, we fit each profile with a model consisting of a (narrow) Gaussian convolved with an exponential scattering tail.
We then use the positions inferred from these fits to shift all pulses to a common phase (multiplying with a phase ramp in the Fourier domain to most easily allow for sub-pixel shifts).
Next, we create summed pulse profiles for each sidereal day, optimally weighting them by normalizing each pulse by the square of the off-pulse rms noise.

It was found, however, that the fitting process is not entirely reliable for a small portion of the individual profiles, with the fits either failing entirely or resulting in visibly incorrect pulse positions.
Furthermore, fitting the same exponential tail model to the resulting average profiles, it becomes clear that this is not always a good model: the scattering tail is often sharper than an exponential tail allows for, falling more quickly near the start and more slowly in the tail.
Better results were obtained with a model consisting of a Gaussian convolved with a ``modified exponential tail'', of the form $\exp((-t/\tau)^{\gamma})$, where $\gamma$ determines the sharpness.
With this added parameter, the observed scattering tails are much more consistently reproduced (see Fig.~\ref{fig:alignment}).
Values of $\gamma$ between 1 and 0.25 were obtained, with lower scattering times typically corresponding to lower $\gamma$, and larger scattering times corresponding to larger $\gamma$.

To align the profiles more reliably, we iterate: we fit the first average we constructed with the improved modified exponential tail model, and then use those fits as a matched filter for the individual pulse profiles of the corresponding sidereal day.
We then obtain improved positions by fitting the resulting peaks with a Gaussian, use those positions to align the profiles, and sum those together as before to construct the final average profiles.

\section{A Cacophony of Echoes}\label{sec:echoes}

% Note used for all panels and the example figure.
% \def \figsetnote[2]{%
\newcommand{\figsetnote}[1]{
  Evolution of the scattering profile.
  {\em Top:\/} Stack of daily average giant pulse profiles for the full data set.
  {\em Middle:\/} Expanded view of the average profiles indicated by the blue box (MJD #1).
  {\em Bottom:\/} Residuals from fits to these average profiles with a narrow Gaussian convolved with a modified exponential tail (shown on a non-linear colour scale to help highlight echoes).
  In the expanded views, the times of closest approach of echoes are marked by green ticks.
  For the more prominent echoes, in the residuals panel parabolic arcs are overlaid, with dashed line style for the side where the echo was visible, and dotted style for the continuation on the other side.}

\figsetstart
    \figsetnum{4}
    \figsettitle{Evolution of the Scattering Profile}

    \figsetgrpstart
    \figsetgrpnum{4.1}
    \figsetgrptitle{Stacked data set}
    \figsetplot{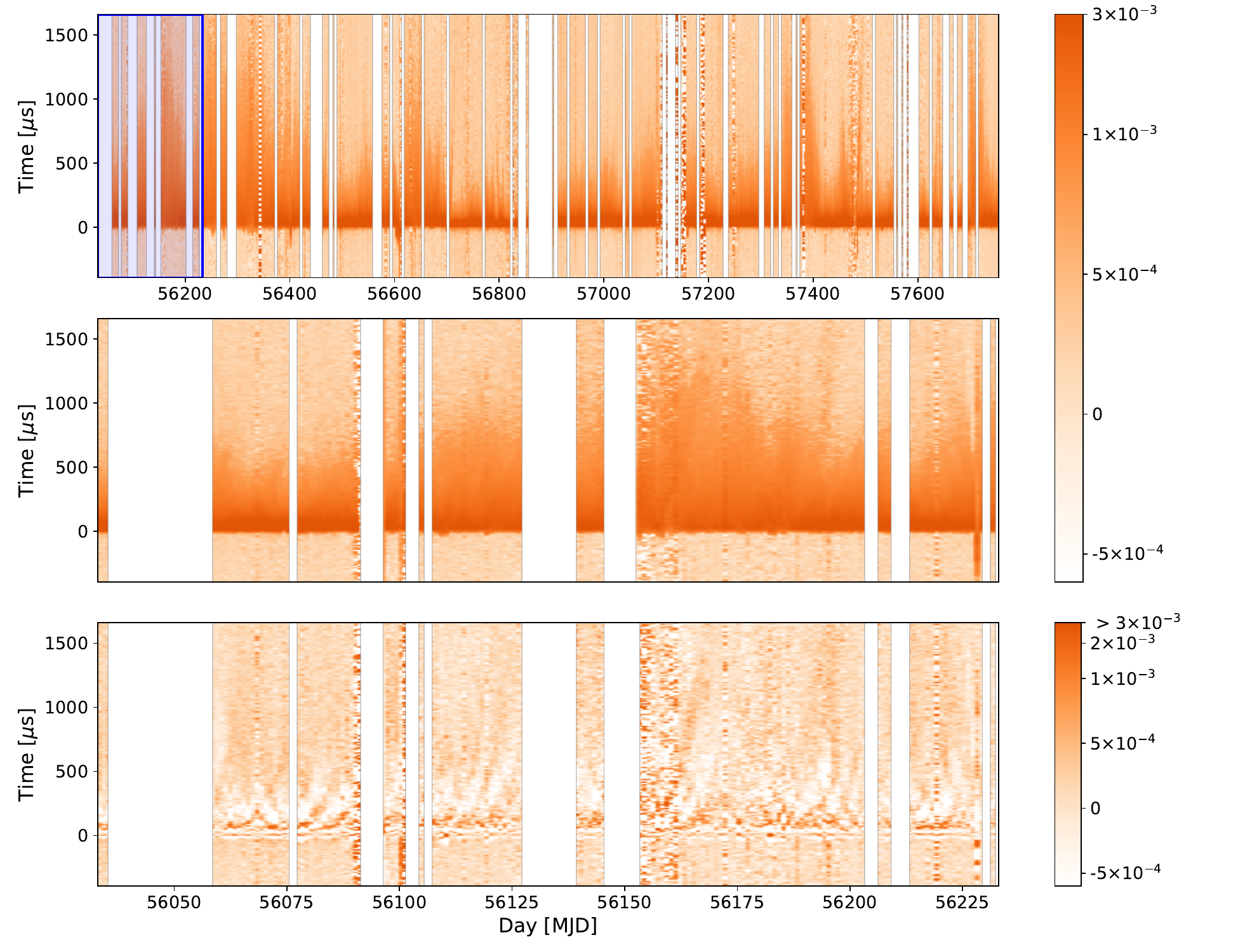}
    \figsetgrpnote{\figsetnote{56033 -- 56233}}
    \figsetgrpend

    \figsetgrpstart
    \figsetgrpnum{4.2}
    \figsetgrptitle{Stacked data set}
    \figsetplot{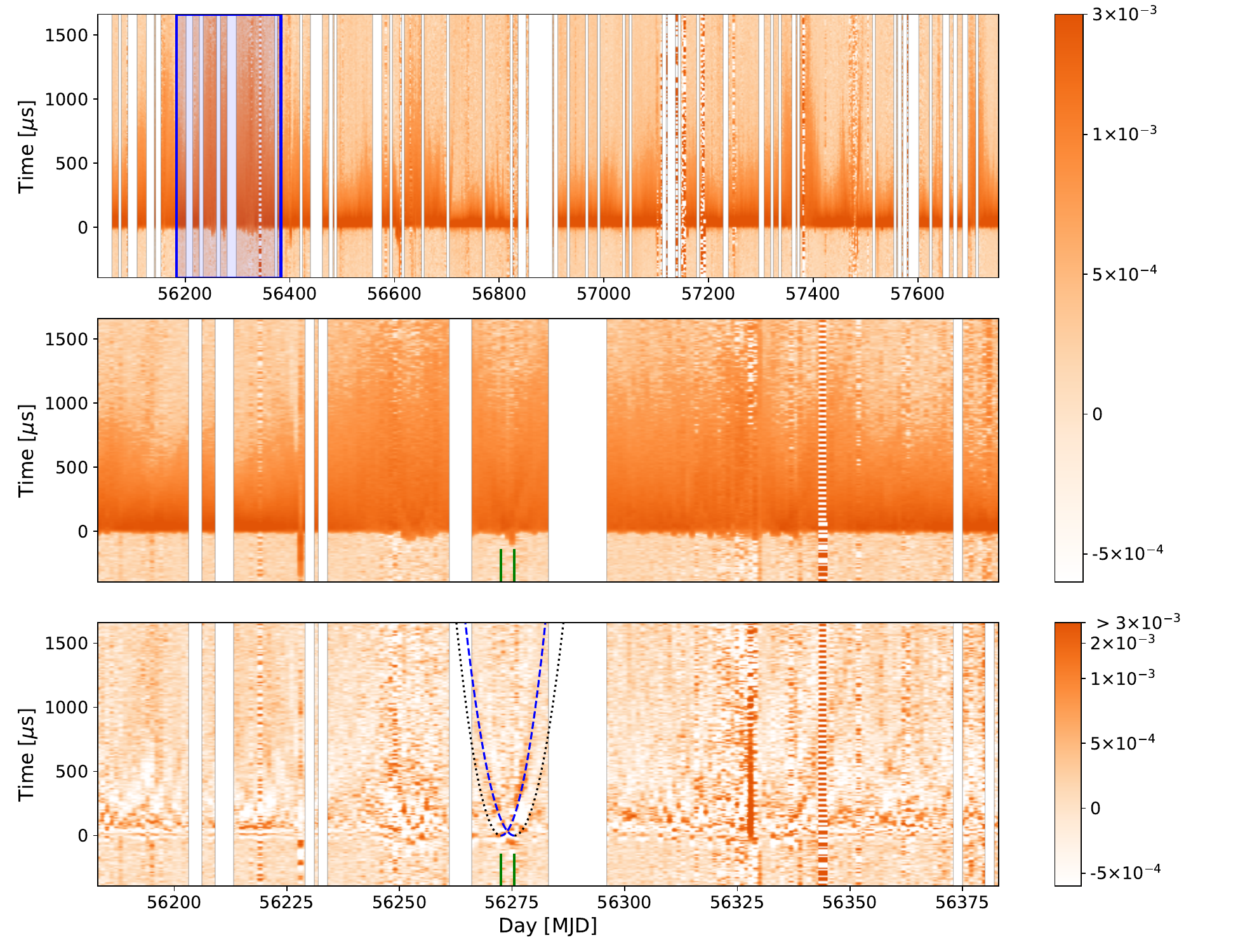}
    \figsetgrpnote{\figsetnote{56183 -- 56382}}
    \figsetgrpend

    \figsetgrpstart
    \figsetgrpnum{4.3}
    \figsetgrptitle{Stacked data set}
    \figsetplot{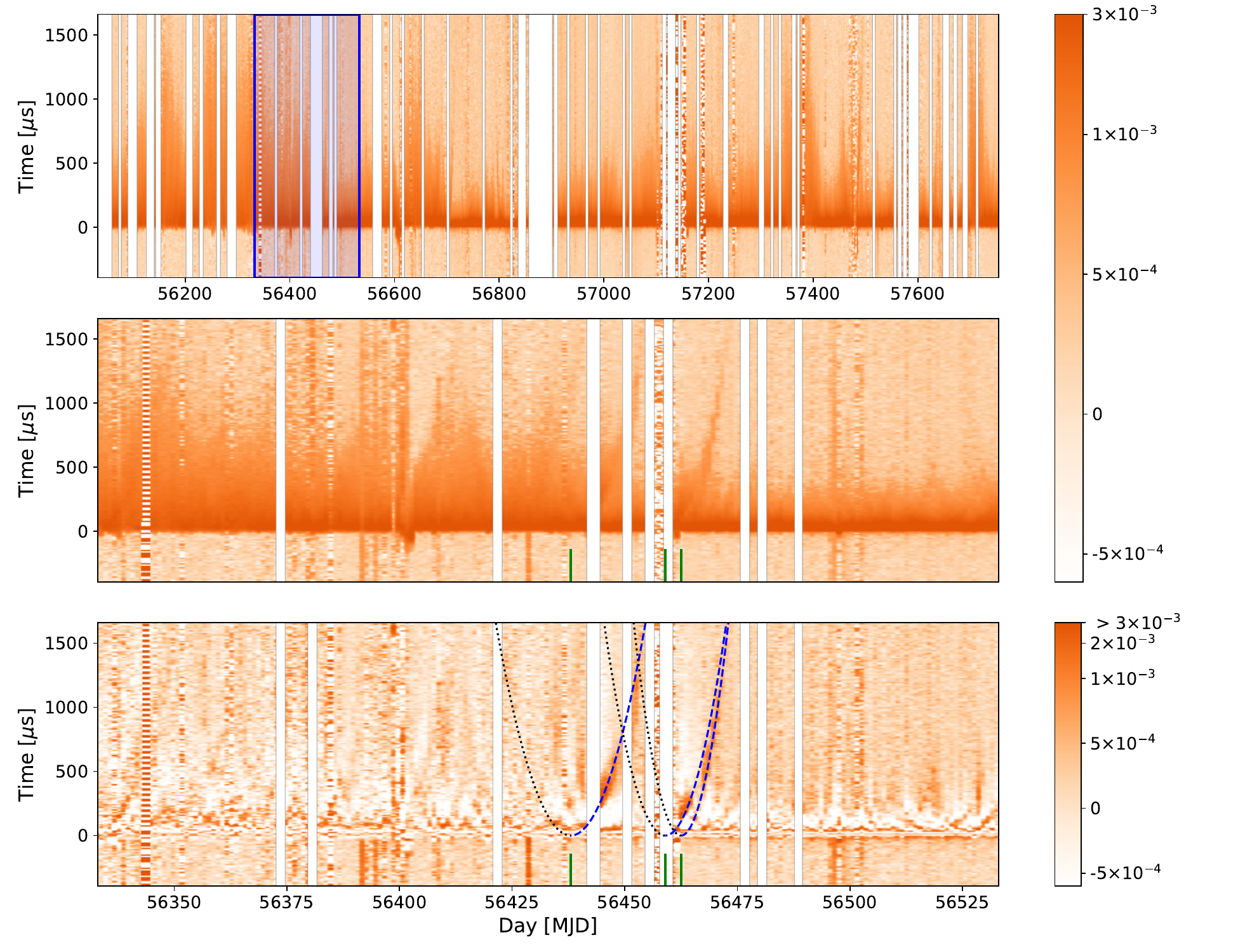}
    \figsetgrpnote{\figsetnote{56333 -- 56532}}
    \figsetgrpend

    \figsetgrpstart
    \figsetgrpnum{4.4}
    \figsetgrptitle{Stacked data set}
    \figsetplot{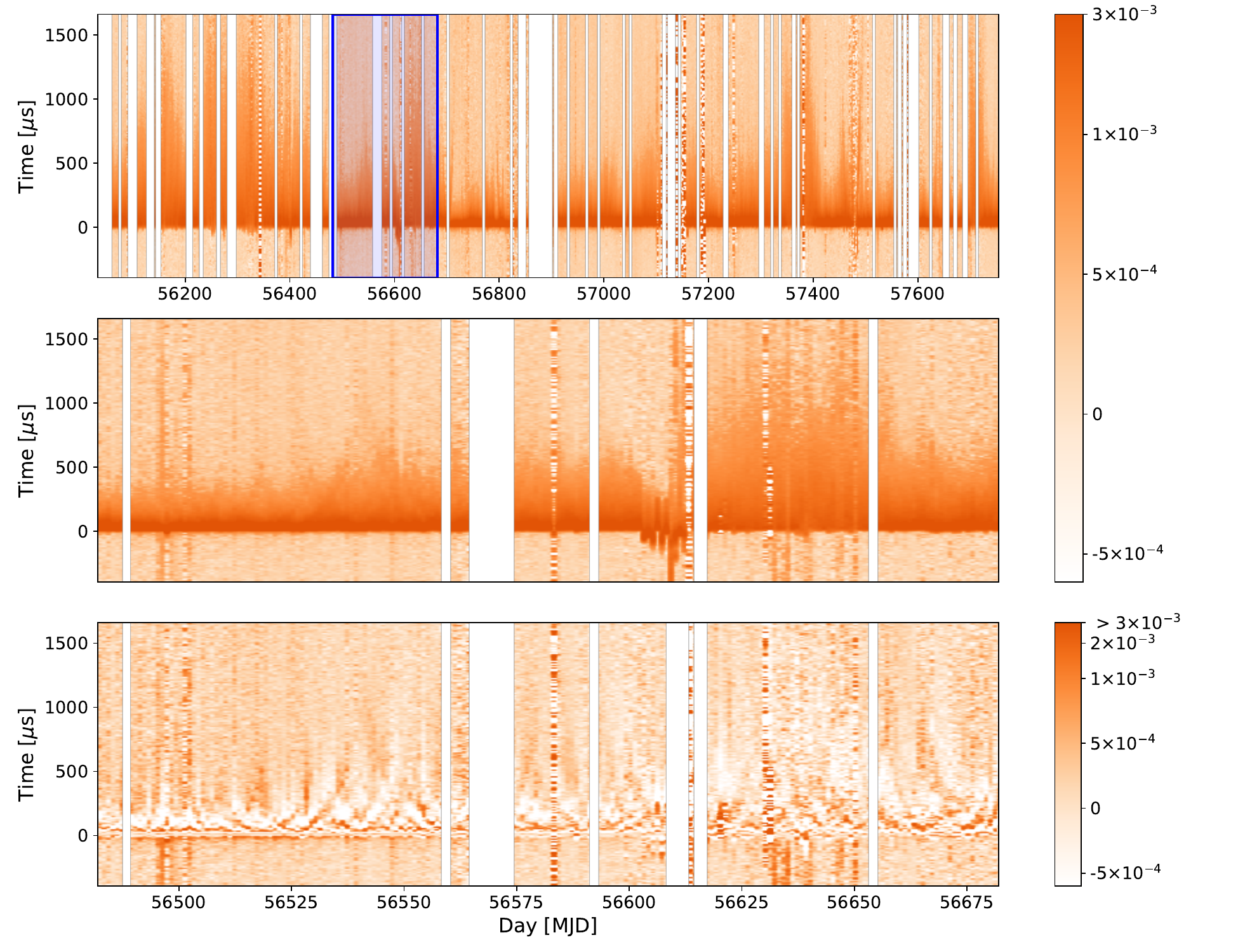}
    \figsetgrpnote{\figsetnote{56482 -- 56682}}
    \figsetgrpend

    \figsetgrpstart
    \figsetgrpnum{4.5}
    \figsetgrptitle{Stacked data set}
    \figsetplot{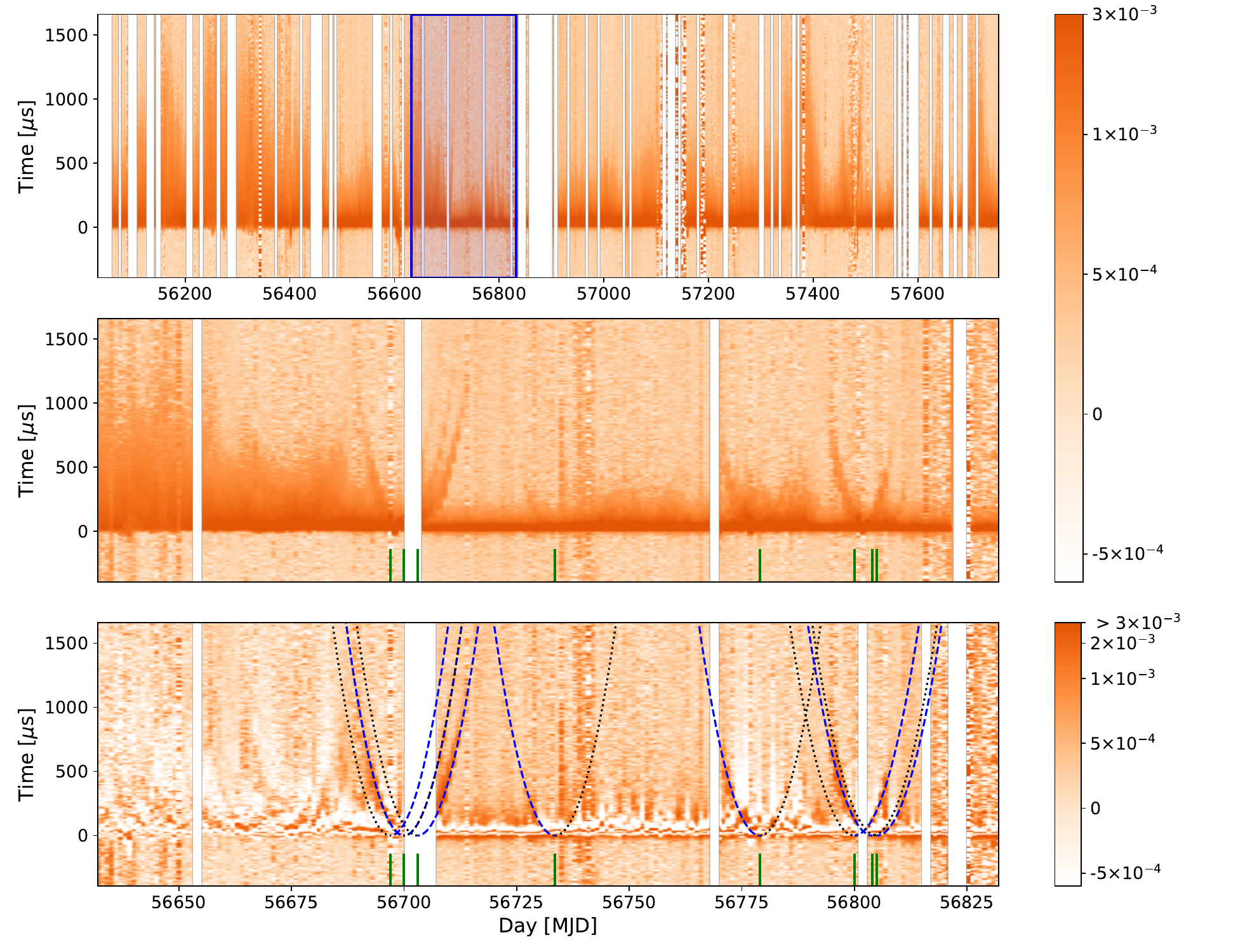}
    \figsetgrpnote{\figsetnote{56632 -- 56831}}
    \figsetgrpend

    \figsetgrpstart
    \figsetgrpnum{4.6}
    \figsetgrptitle{Stacked data set}
    \figsetplot{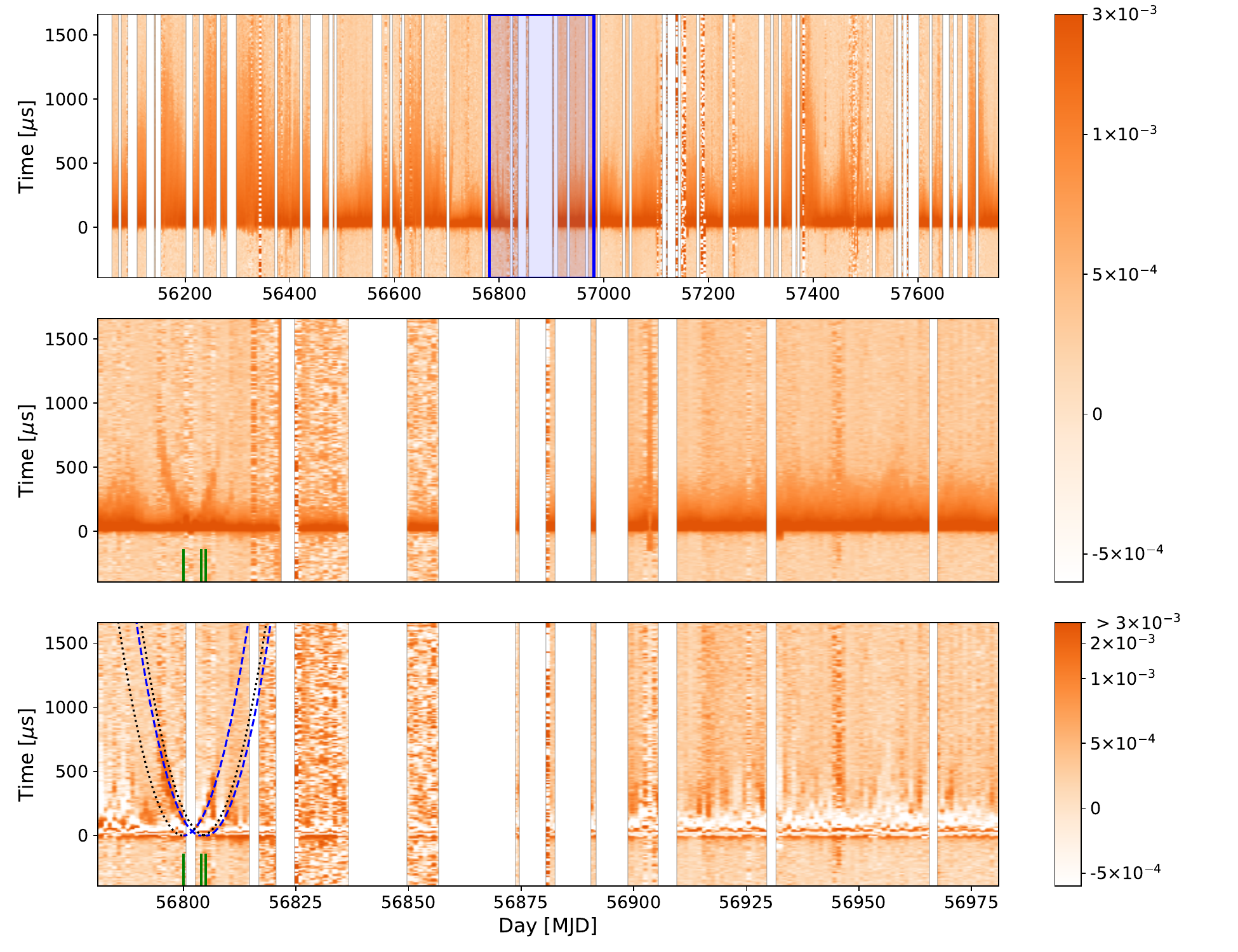}
    \figsetgrpnote{\figsetnote{56781 -- 56981}}
    \figsetgrpend

    \figsetgrpstart
    \figsetgrpnum{4.7}
    \figsetgrptitle{Stacked data set}
    \figsetplot{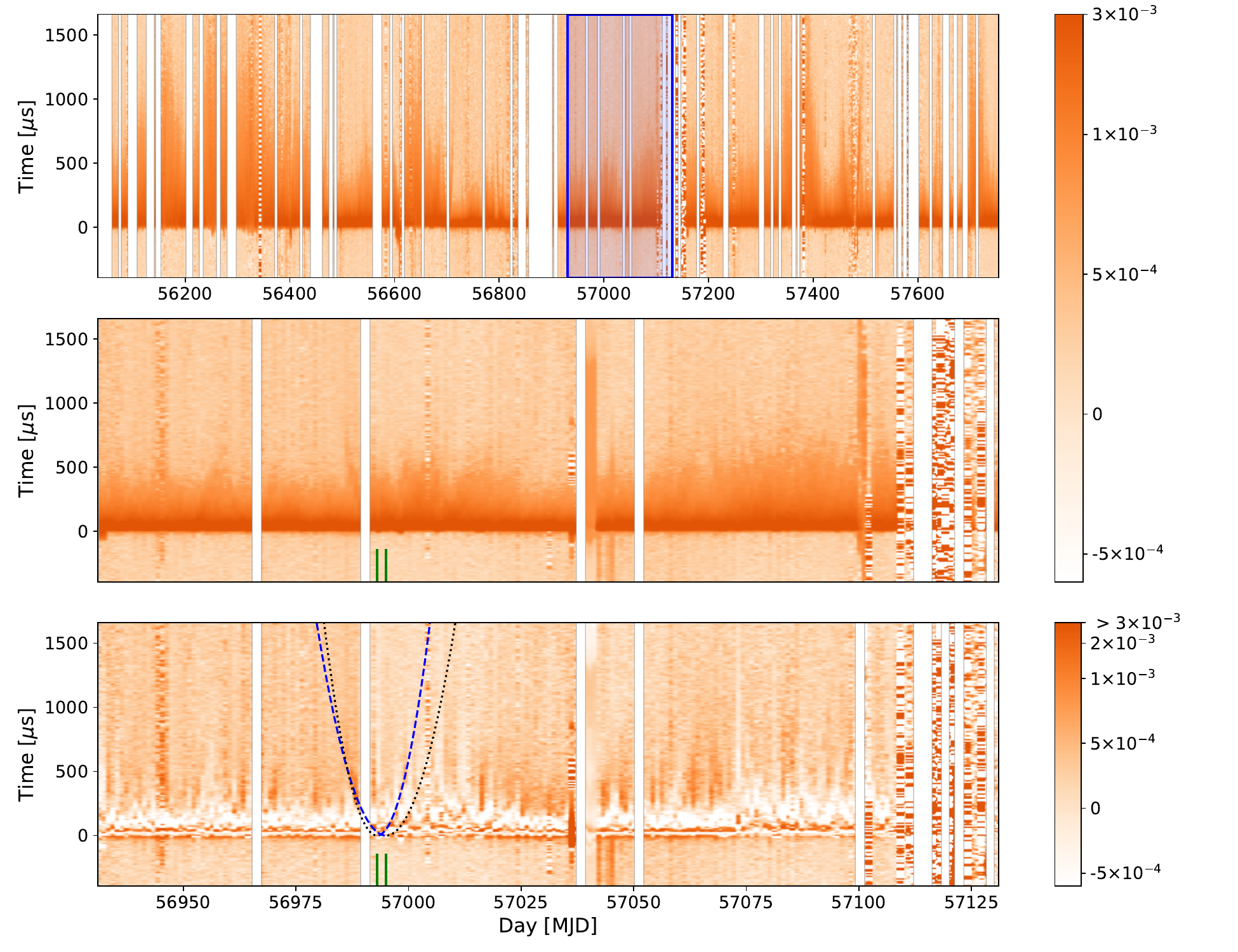}
    \figsetgrpnote{\figsetnote{56931 -- 57130}}
    \figsetgrpend

    \figsetgrpstart
    \figsetgrpnum{4.8}
    \figsetgrptitle{Stacked data set}
    \figsetplot{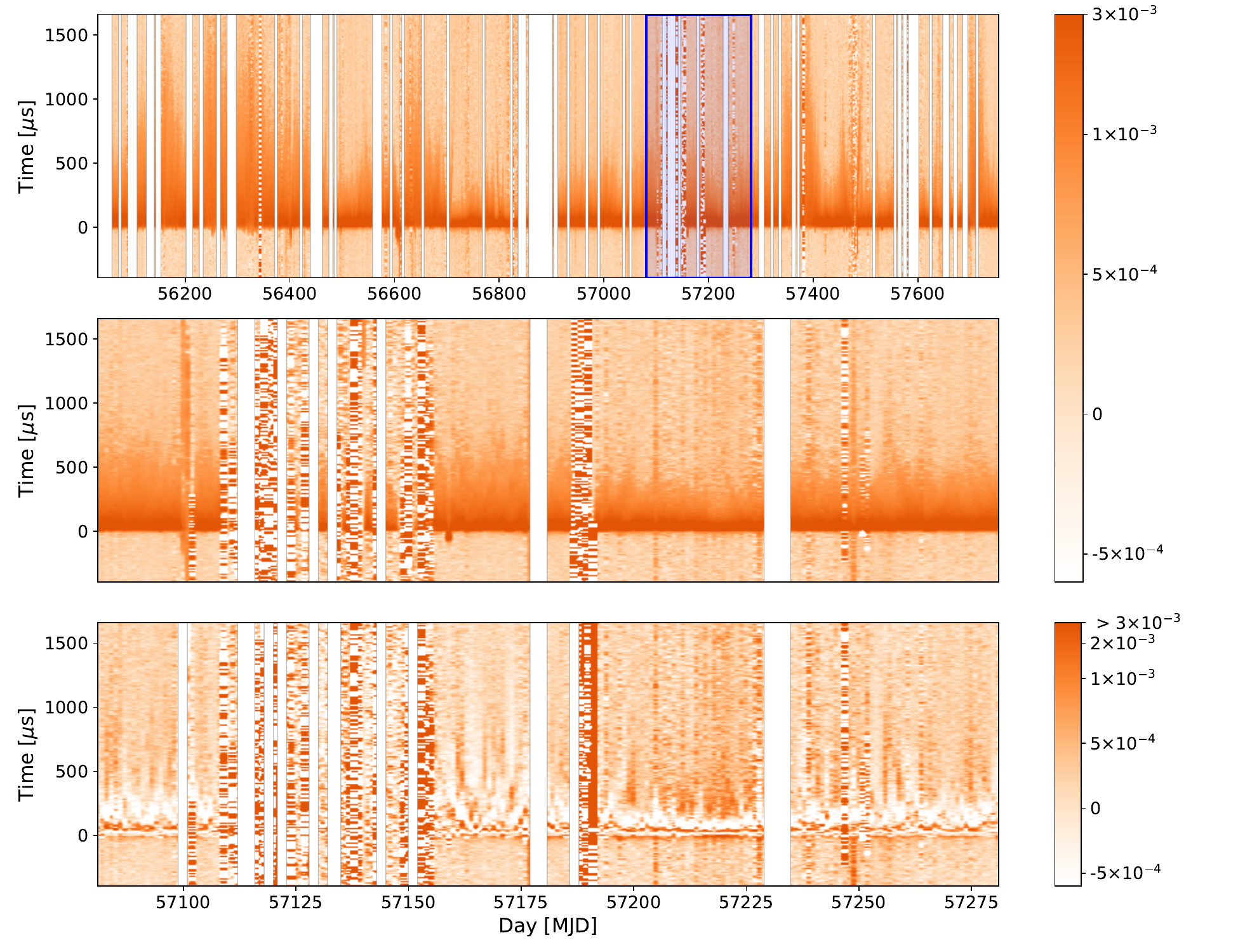}
    \figsetgrpnote{\figsetnote{57081 -- 57280}}
    \figsetgrpend

    \figsetgrpstart
    \figsetgrpnum{4.9}
    \figsetgrptitle{Stacked data set}
    \figsetplot{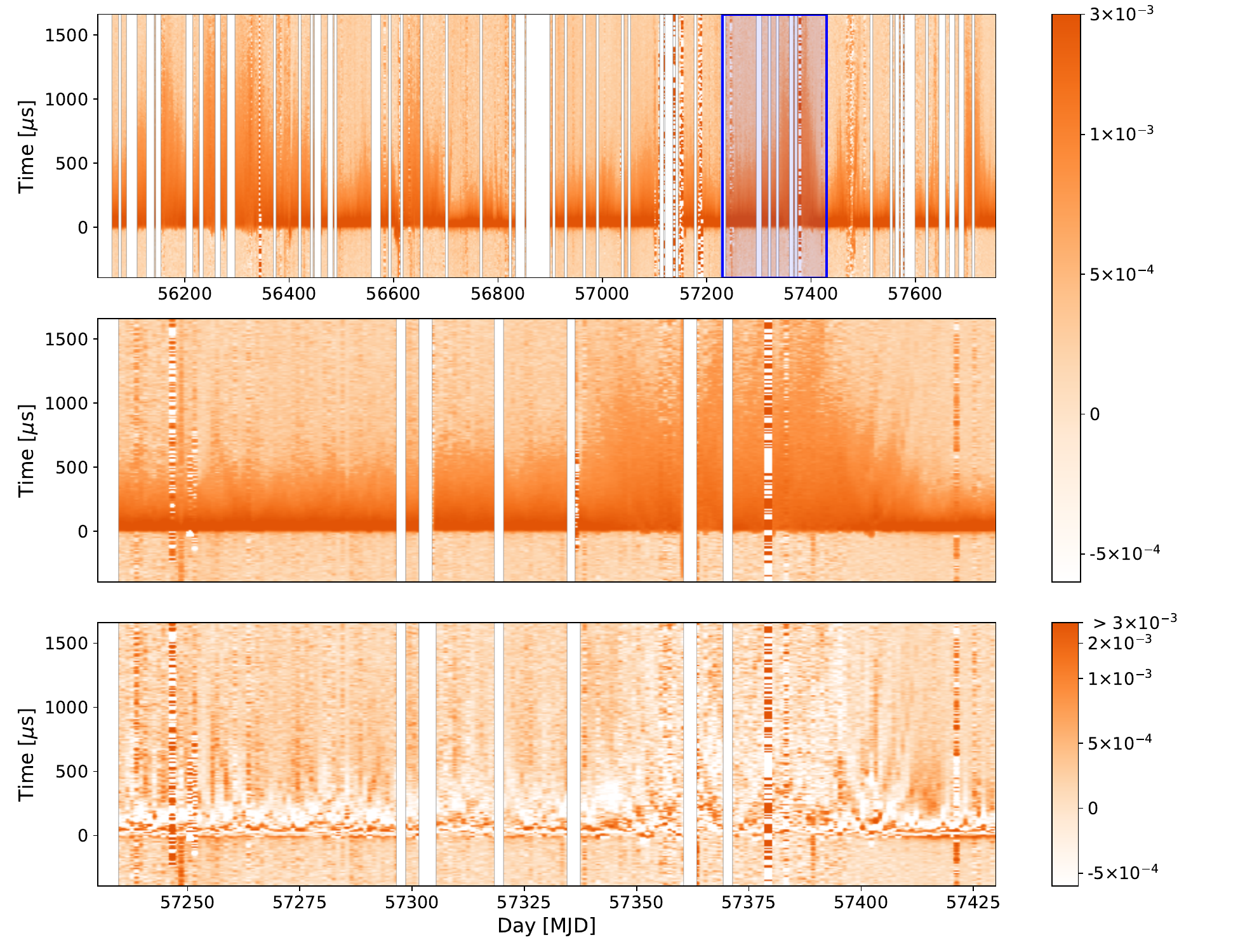}
    \figsetgrpnote{\figsetnote{57230 -- 57430}}
    \figsetgrpend

    \figsetgrpstart
    \figsetgrpnum{4.10}
    \figsetgrptitle{Stacked data set}
    \figsetplot{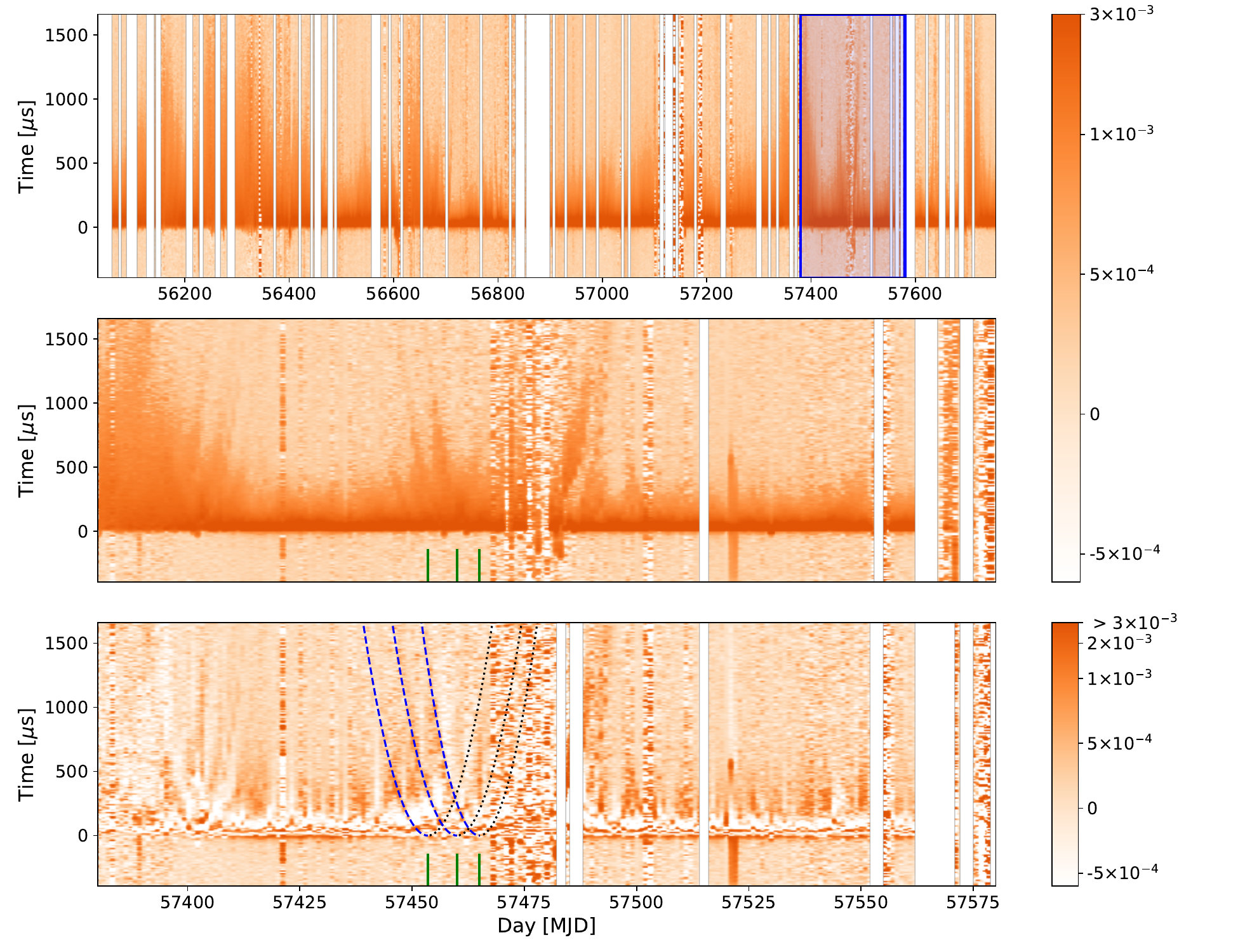}
    \figsetgrpnote{\figsetnote{57380 -- 57579}}
    \figsetgrpend

    \figsetgrpstart
    \figsetgrpnum{4.11}
    \figsetgrptitle{Stacked data set}
    \figsetplot{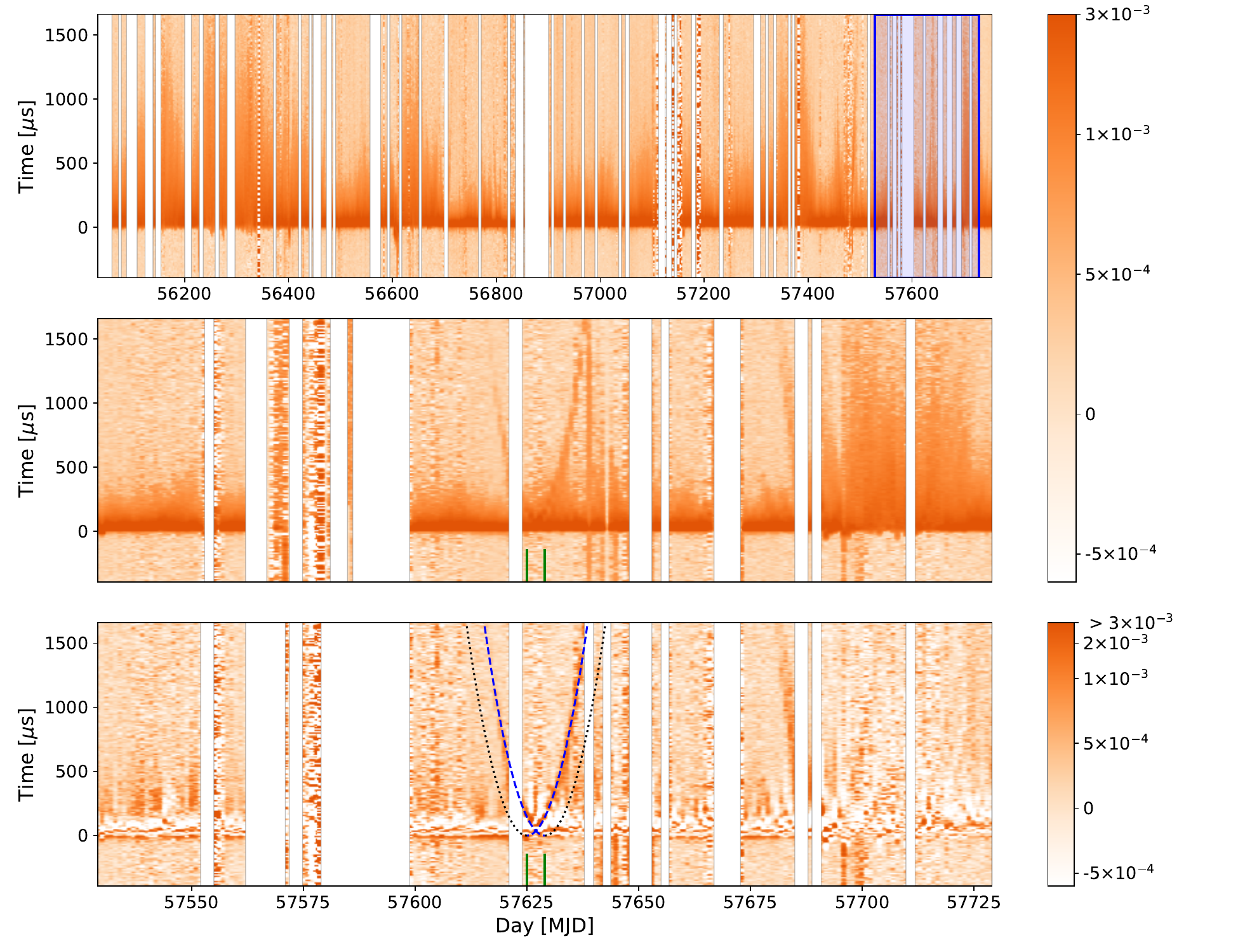}
    \figsetgrpnote{\figsetnote{57529 -- 57729}}
    \figsetgrpend

    \figsetgrpstart
    \figsetgrpnum{4.12}
    \figsetgrptitle{Stacked data set}
    \figsetplot{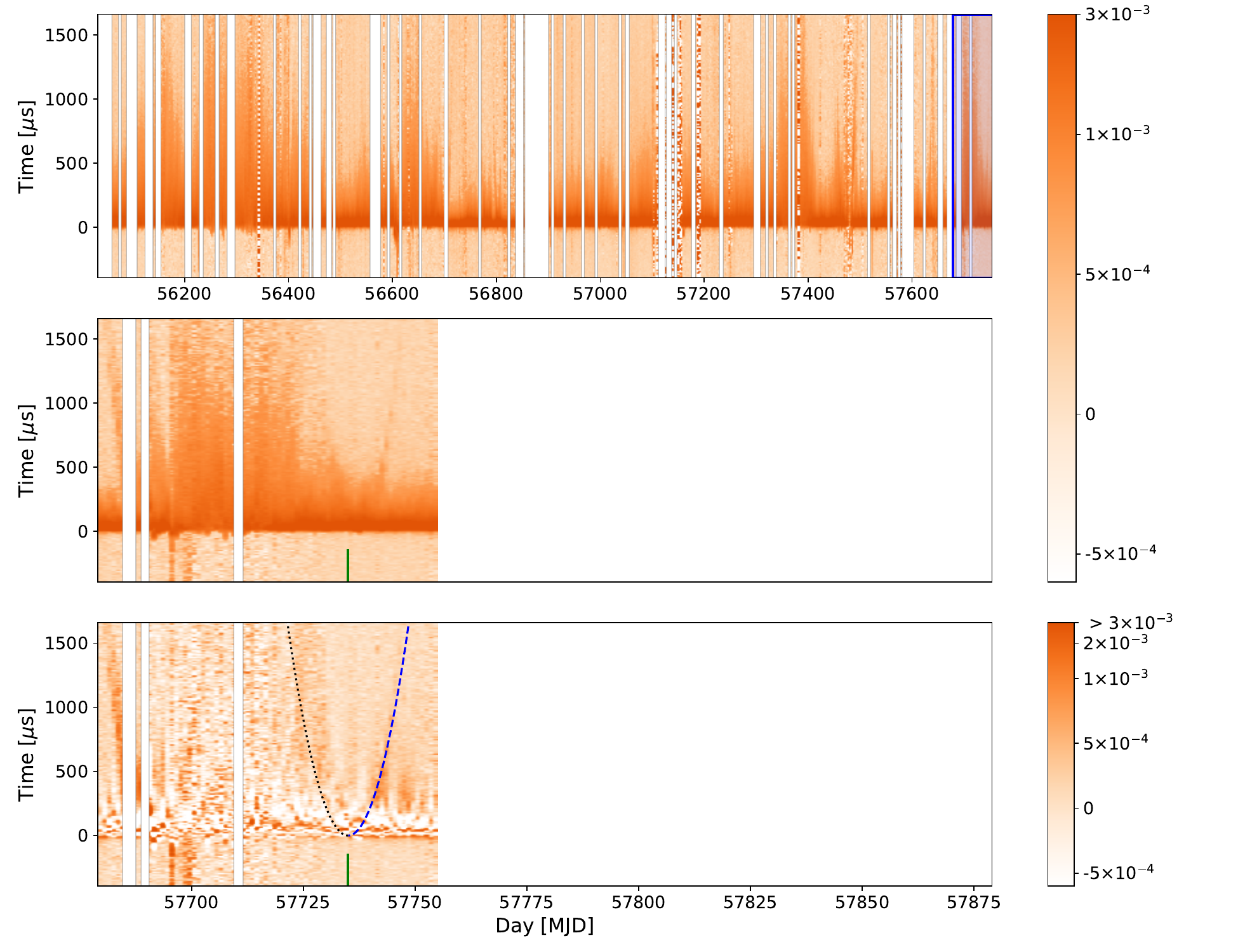}
    \figsetgrpnote{\figsetnote{57679 -- 57879}}
    \figsetgrpend

\figsetend

\begin{figure*}
  \centering
  \includegraphics[width=\textwidth]{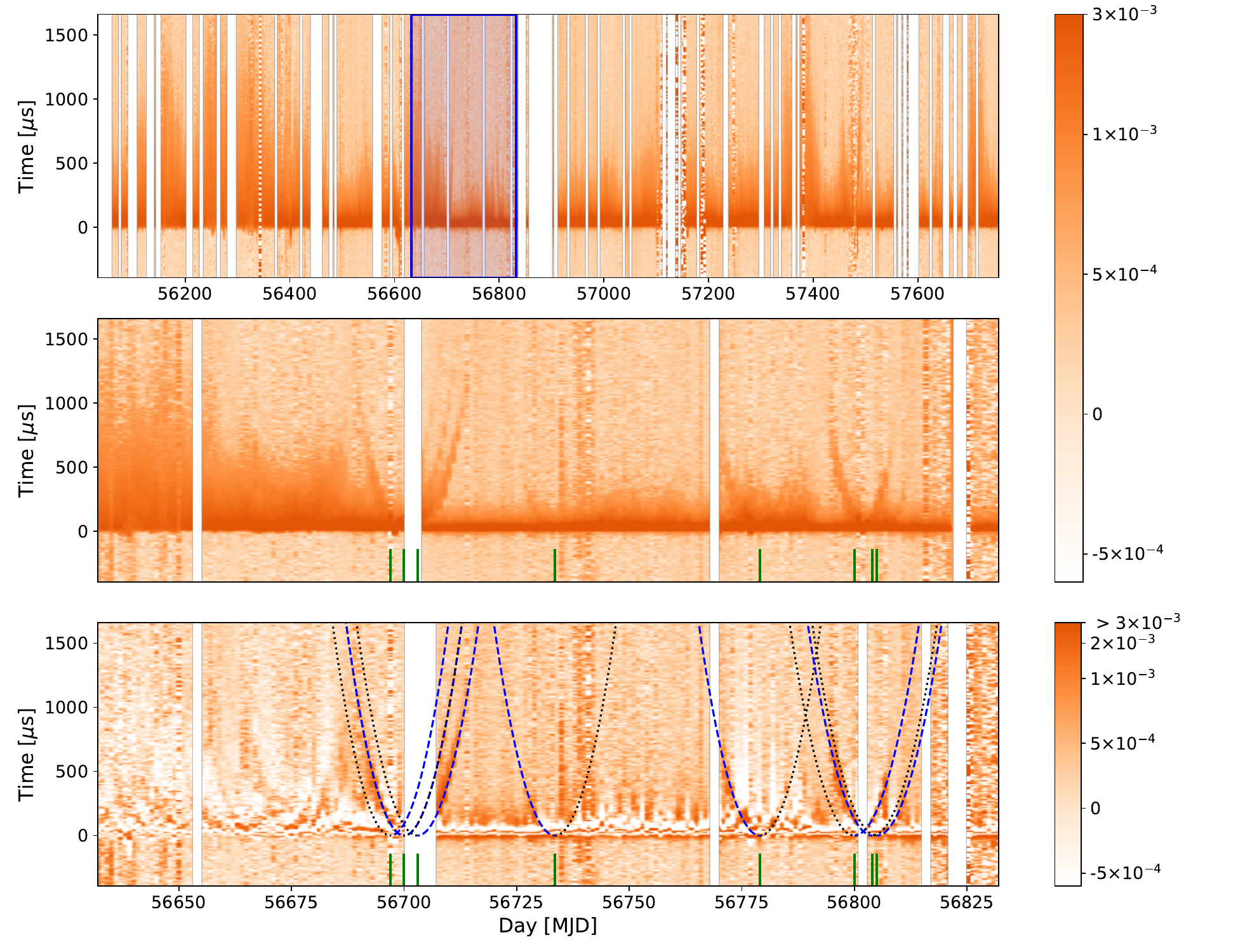}
  \caption{\figsetnote{56632 -- 56831}\
    The complete figure set (12 figures) is available in the online journal.
    \label{fig:echoes}\label{fig:yearplot}
  }
\end{figure*}

The daily average profiles show the strong scattering variations expected from the Crab's scattering environment, as is immediately obvious in Figure~\ref{fig:yearplot}, where they are combined into a stack.
Also visible in the stack, especially when zoomed in to a finer timescale, are prominent echoes, some with incoming and some with outgoing arcs.

Inspecting the data set, a number of things stand out.
First, strong echoes often appear clustered together, suggesting a possible causal connection, perhaps to some nebular structure.

Second, the clearest echoes appear during days with lower scattering.
This might be a signal-to-noise issue: fewer giant pulses are detected on days with higher scattering, and echoes are more difficult to see against a longer scattering tail.
Given the observed clustering of prominent echoes, however, it may also be that observed periods of high scattering are at least partly the result of the presence of many echoes, which blend together to give the appearance of a longer scattering tail.

Third, during periods of low scattering, many less prominent echoes appear to be present.
Unlike the clearer cases discussed above, these are difficult to see directly in the stacks but seem convincingly identified as a repeating pattern of arclet structures in the stacks of the residuals between the daily average profiles and the fitted scattering tail model (see Fig.~\ref{fig:echoes}).

Fourth, as previously noted by \citet{lyne01}, the observed echoes trace arcs that are consistent with approaching zero delay, implying that the structures responsible closely approach or directly cross the line of sight to the Crab.
This is unexpected if the structures responsible are some kind of ``blobs,''  as for any roughly spherical, localized structures directly crossing the line of sight should be the exception rather than the rule, i.e., the typical observed minimum delay of the echoes should be nonzero.
Instead, it suggests structures that are highly anisotropic (as projected on the sky), causing an echo that moves along their long axis, such that as the pulsar moves, the echo will be on the line of sight when it crosses the structure.

Fifth, the arcs look parabolic, with the delay depending quadratically on time relative to the zero-delay crossing.
This is expected generically for linear motion relative to a fixed scattering structure, with the curvature depending on a combination of distance, orientation, and relative velocity (see Sect.~\ref{sec:rams}).
We matched parabolae to the most prominent echoes in our data set (see Fig.~\ref{fig:echoes}) and found that the curvatures are not the same, but vary between $6$ and $17{\rm\,\mu s/day^2}$, implying differences in distances, orientations or velocities.
We also find that while echoes clustered together in time have similar curvatures, within $2{\rm\,\mu s/day^2}$, they are not identical.
Thus, if these are caused by the same larger-scale structure and thus have the same distance and velocity, their relative orientations must differ.

Sixth, even the most prominent echoes only rarely, if ever, seem to continue across zero (where they have parts on both sides with the same curvature).
This suggests that a given structure can typically bend radiation only in one direction.

Overall, the results show that echoes are more common than previously realized, and will be apparent when care is taken to correct for jitter in individual pulses before stacking.
They also suggests that a ``cacophony of echoes'' is always present and that echoes may at least be partially responsible for the extent of scattering tails.
Furthermore, the structures responsible are likely highly elongated (as projected on the sky) and can bend light in only one direction.

\section{Properties of the Echo Structures}\label{sec:structures}

The scattering is thought to occur in the Crab nebula, and is likely related to the filaments that are seen in optical emission lines.
Given that those filaments are highly elongated, it is not unexpected that structures associated with them on smaller scales, including those that could cause the echoes we see, are elongated as well.
Below, we check whether our observations are consistent with this picture, first discussing constraints on the locations of the scattering structures, and then turning to their shapes.

\subsection{Locations}\label{subsec:location}

An elongated structure in the Crab nebula (i.e., much closer to the pulsar than to us), which is crossed by the pulsar's projected trajectory,  will produce an echo with a geometric delay $\tau$ that varies quadratically with time as,
\begin{equation}
  \tau(t) = \delta(t)^2\frac{d_{ps} }{2c}
  = \frac{\left( v_{\rm eff}\cos\psi\right)^2}{2cd_{ps}}t^2 \equiv \eta t^2,
  \label{eqn:delay}
\end{equation}
where $\delta(t)$ is the angular offset as a function of time $t$ (relative to the crossing time), $d_{ps}$ the distance between the pulsar and the screen, and $\psi$ is the angle between the position angle of the normal to the structure and the effective velocity $v_{\rm eff}$, and where in the final part we implicitly defined the curvature $\eta$.
The effective velocity is given by the difference between the pulsar and structure velocities projected onto the sky,
\begin{equation}\label{eqn:velocity}
  \vec{v}_{\rm eff} = \vec{v}_{p, \rm sky} - \vec{v}_{s, \rm sky}.
\end{equation}
Note that these are just the normal equations for scattering of a thin screen (e.g., \citealt{cordes98}), but with the simplification that the screen is close to the pulsar.

\begin{figure}
  \centering
  \includegraphics[width=\hsize]{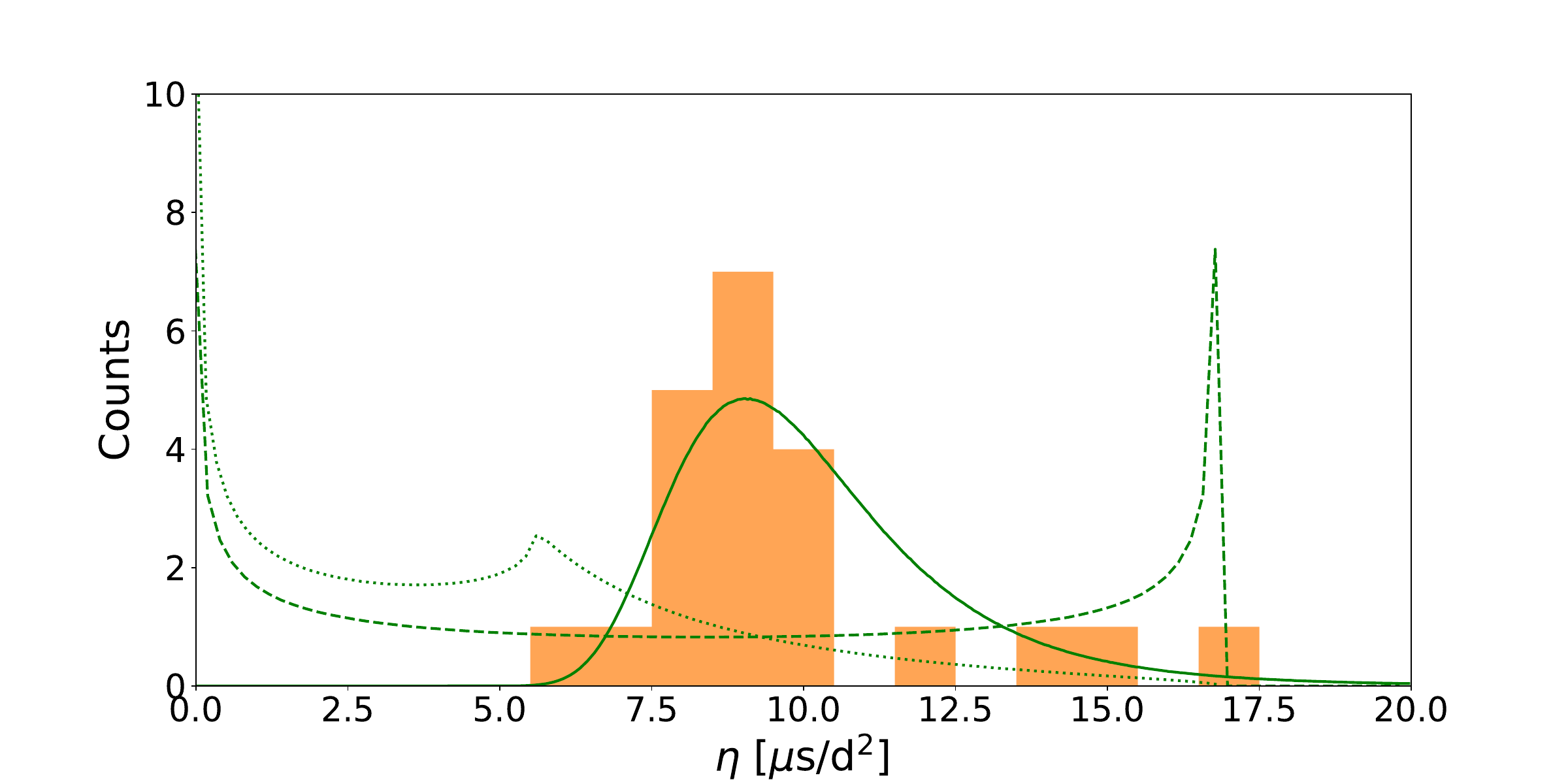}
  \caption{
    Distribution of the measured curvatures for the most prominent echoes.
    Overlaid are curves expected to be generated by structures that are located anywhere in the nebula with random orientations ($d_{ps}\sim{\cal U}(0.5, 2.0){\rm\,pc}$; dotted line), at the edge of the wind nebula with random orientations ($d_{ps}=0.5{\rm\,pc}$; dashed line), and near the edge of the nebula with a small range of orientations ($d_{ps}\sim{\cal N}(0.5, 0.1){\rm\,pc}$, $\psi\sim{\cal N}(40.7, 0.6){\rm\,deg}$; full line).
    \label{fig:etas}
  }
\end{figure}

The distribution of the curvatures $\eta$ for the observed echoes is shown in Fig.~\ref{fig:etas}.
Most curvatures lie in the range between $\eta=$6 and $\eta=10{\rm\,\mu s/day^2}$, while a few have higher values, yielding an average curvature of $\bar{\eta}=9.7{\rm\,\mu s/day^2}$.

For a given echo arc, the curvature $\eta$ constrains a combination of the corresponding structure's distance from the pulsar, its effective velocity, and its orientation, with the highest curvatures providing the most stringent constraints.
From observations of nebular emission, \cite{martin21} found that the optical filaments roughly extend from 0.5 to $2.0{\rm\,pc}$ (where the inner boundary corresponds to the edge of the pulsar wind nebula; \citealt{hess20}).
If the material responsible for producing the fastest echoes, with $\eta=17{\rm\,\mu s/day^2}$ is located at least $0.5{\rm\,pc}$ from the pulsar, it requires that $v_{\rm eff} \gtrsim 145\,\textrm{km/s}$.
This is slightly higher than the pulsar's velocity of $v_p=120\pm 23{\rm\,km/s}$ inferred from astrometry \citep{kaplan08}, though still consistent given the relatively large uncertainties on the pulsar velocity relative to the center of the nebula and the possible contribution to the effective velocity of the motion of nebular material.

Scaling $v_{\rm eff}$ to this velocity, an estimate of the distance to any echo is then,
\begin{equation}\label{eqn:distance}
d_{ps} = 0.85{\rm\;pc} \left( \frac{\eta}{10{\rm\,\mu s/day^2}} \right)^{-1}
\left( \frac{v_{\rm eff} \cos\psi}{145{\rm\,km/s}} \right)^2,
\end{equation}
which is an upper limit, since $\cos\psi\leq1$.
From our observed curvature range of $6 \leq \eta \leq 17{\rm\,\mu s/day^2}$, we therefore infer a range in maximum distances of 0.5 to $1.4{\rm\,pc}$.
Here, the lower bound matches the \cite{martin21} range by construction, but the fact that the upper boundary is also within the expected range from the nebular emission gives some confidence in the estimates.

The distribution in curvature gives further information, as can be seen in Fig.~\ref{fig:etas}, where we compare it with expectations for three distinct cases.
The first two assume random screen orientations, with structures either placed uniformly between 0.5 and $1.5{\rm\,pc}$ or at a fixed distance of $0.5{\rm\,pc}$.
Neither matches, as they predict a peak at very low curvature that is not seen (corresponding to  $\cos\psi\simeq0$, i.e., structures nearly aligned with the pulsar motion so that the separation between the structure and the line of sight varies slowly).

In the third case, we sample both screen distance and orientation from narrow Gaussian distributions.
This case reproduces the observed distribution of echo curvatures much better.
As expected from Eq.~\ref{eqn:distance}, there is a degeneracy between the mean distance and mean orientation: given a minimum mean screen distance $\bar{d}_{ps}>0.5{\rm\,pc}$ and assuming $v_{\rm eff}=145{\rm\,km/s}$, we constrain the mean orientation to $\bar{\psi} \leq 40.9\arcdeg$; conversely, given that $\bar\psi\geq0\arcdeg$, we constrain the mean distance to $\bar{d}_{ps} \leq 0.87{\rm\,pc}$.

Regardless of the precise distance or orientation, the range in both $d_{ps}$ and $\psi$ has to be relatively small.
We return to possible interpretations of this in Sect.~\ref{sec:rams}.

\begin{figure*}
  \centering
  \includegraphics[width=0.8\textwidth]{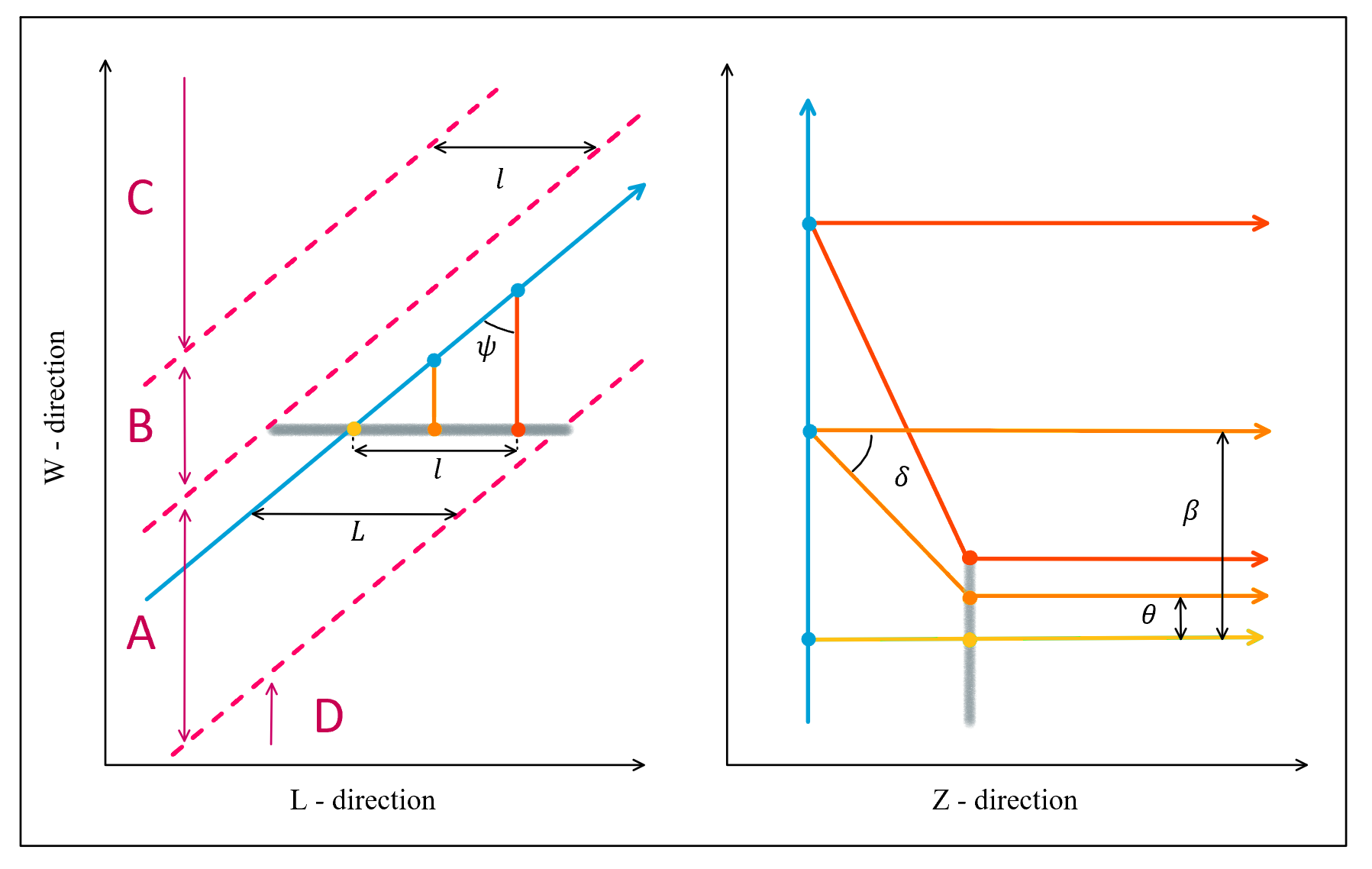}
  \caption{Geometry of an echo, with an on-sky view on the left, and one along the line of sight on the right.
    The grey line is the structure responsible for the echoes (which can only bend light in one direction), and the blue arrow marks the pulsar's path.
    The rays of the lensed image are shown at three different times (yellow, orange and red), with the corresponding line-of-sight paths also shown on the right.
    In the left-hand panel, four regions are defined:
    (A) Echoes will be seen starting at zero delay until the maximum distance for which the structure can still bend light towards the line of sight;
    (B) Echos will be seen starting at non-zero delay;
    (C) The pulsar passes at a distance too far from the structure to see echoes; and
    (D) The pulsar passes on the wrong side of the structure to see echoes.
    Note that the right-hand panel is strongly compressed along the line of sight and expanded along the width ($W$) relative to the left-hand panel;
    we infer from our observations that the width (vertical extent here) is far smaller than the depth $Z$ (horizontal extent).
    The orientation angle of the screen relative to the pulsar motion is $\psi$, $\delta$ is the bending angle, and $\theta$ and $\beta$ are the angular offsets between the structure and the echo and line of sight images, respectively, as seen by the observer.
    \label{fig:sketch_los}}
\end{figure*}

\begin{figure*}
  \centering
  \includegraphics[width=0.8\textwidth]{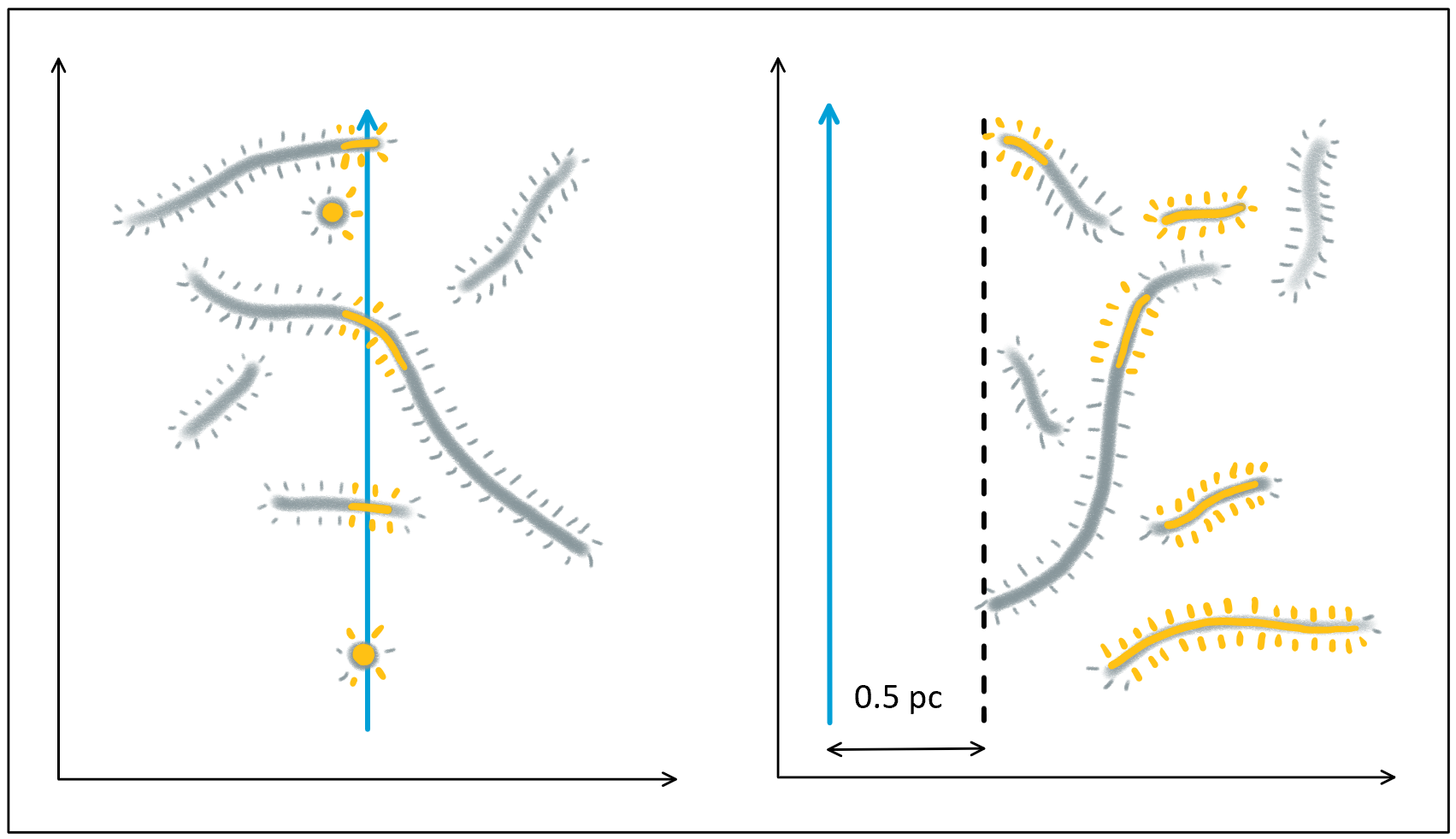}
  \caption{Sketch of the large-scale geometry of the structures causing echoes, with an on-sky view on the left, and one along the line of sight on the right.
    Like in Figure~\ref{fig:sketch_los}, the blue arrow is the path of the pulsar.
    Filaments are drawn as thick grey lines, with regions that may exhibit echoes highlighted in yellow.
    Short ``hairs'' along the filaments represent the ripples or smaller-scale Rayleigh-Taylor instabilities which we suggest may be directly responsible for the echoes.
    \label{fig:sketch_filaments}}
\end{figure*}

\subsection{Shapes}\label{subsec:prop}

The strengths and durations of echoes are influenced by the typical electron densities $n_e$ of the structures, as well as their size and shape.
Hence, we can use the echoes and estimates of $n_e$ from optical emission lines to constrain the typical length $L$, width $W$ and depth $Z$ of the lensing structures.

The first qualitative constraint we have is that the structures must be highly elongated, i.e., $L \gg W$, in order for the echoes to last long and have a high likelihood of approaching a minimum delay of zero (see Sect.~\ref{sec:echoes}).
As the lensed image moves it will be on the part of the structure closest to the pulsar, moving along with it (see Fig.~\ref{fig:sketch_los}).
Thus for any observed echo, one can estimate a minimum size $\ell_i$ that is needed to ensure the structure is visible for a given duration $\Delta t_i$:
\begin{eqnarray}
  \ell_i &=& v_{\rm eff}\Delta t_i \sin|\psi_i|\nonumber\\
  &=& 0.8{\rm\,AU}\,\sin|\psi_i|
  \left(\frac{\Delta t_i}{10 \textrm{days}}\right)
  \left(\frac{v_{\rm eff}}{145{\rm\,km/s}}\right).
\end{eqnarray}
Furthermore, for a given structure with total length $L_i$, the probability that the resulting echo will approach a minimum delay $\tau_{\rm min} = 0{\rm\,\mu s}$ is the same as the probability that the line of sight will intersect the structure as the pulsar moves on the sky (see Fig.~\ref{fig:sketch_los}), giving $P_{{\rm LOS}, i} = L_i / (L_i + \ell_i)$.
Under the simplifying assumption that these structures share a typical length $L_i \sim L$ and that $L \gg \ell_i$, the probability that out of N observed echoes, all have $\tau_{\rm min} = 0{\rm\,\mu s}$ is then,
\begin{equation}
  P_{LOS}(N) = \prod_{i=1}^{N} \frac{L}{L+\ell_i} \simeq \left( \frac{L}{L + \langle \ell \rangle} \right)^N
\end{equation}
where $\langle \ell \rangle$ is some suitable average of the estimates for individual echoes.

In order to ensure that such an occurrence is not too rare, we require $P_{LOS}(N) \geq P_c$, where $P_c$ is the desired (small) probability that seeing only zero-crossing echoes was a fluke. Rearranging, this corresponds to a lower limit to the length,
\begin{equation}
  L \geq \frac{P_c^{1/N}}{1 - P_c^{1/N}} \langle \ell \rangle.
\end{equation}
Given that in our $N = 22$ observed echoes with measured curvatures, none have non-zero minimum delays, we infer that with 90\% confidence (i.e., $P_c = 0.1$),
\begin{equation}
  L \geq 4{\rm\,AU}\,
  \left(\frac{\sin(\bar{\psi})}{0.5}\right)
  \left(\frac{\Delta t}{10 \textrm{days}}\right)
  \left(\frac{v_\textrm{eff}}{145{\rm\,km/s}}\right).
\end{equation}

To constrain the width W and depth Z of the structures, we need to make an assumption about how they create echoes.
For the width, we will assume that they act like lenses, and for the depth that light is bent by gradients in electron column density.

By conservation of surface brightness, generally the expected magnification for a lens is $\mu = {\rm d}\theta/{\rm d}\beta$, where $\theta$ and $\beta$ are the echo and source positions relative to the structure, respectively.
The position $\theta$ is constrained to be within the lens, being offset by a small amount when the echo is bright and approaching the angular half-width $\omega\equiv W/2d_{ps}$ as it becomes faint.
Hence, for a faint image, one has $\mu\simeq\omega/\beta$.
Furthermore, for a faint image, $\omega\ll\beta$ and hence the angular offset between the source and the echo image $\delta=\beta-\theta\simeq\beta$.

Given the above, we can estimate the physical width of the structure with,
\begin{equation}\label{eqn:X}
  W \simeq 2 \mu \delta d_{ps}
  \simeq 0.1{\rm\;AU}\left(\frac{\mu}{0.05}\right)
  \left(\frac{\delta}{1\arcsec}\right)
  \left(\frac{d_{ps}}{1{\rm\;pc}}\right),
\end{equation}
where we scaled $\delta$ to a typical value (from Eq.~\ref{eqn:delay} using $\tau\simeq1{\rm\,ms}$ and $d_{ps}\simeq1{\rm\,pc}$).

To constrain the depth $Z$, we use that for a given bending angle -- in our case the same as the angular offset $\delta$ implied by an echo's delay $\tau$ -- the required gradient in electron column density $N_e$ is,
\begin{equation}
  \nabla_x N_e = \frac{2\pi \delta}{\lambda^2r_e}
  \simeq 4.5\times10^4{\rm\;cm^{-2}\,cm^{-1}}\left(\frac{\delta}{1\arcsec}\right),
\end{equation}
where $x$ is the direction along which images form, perpendicular to the structure, $\lambda\simeq0.49{\rm\,m}$ is the observing wavelength, and $r_e$ is the classical electron radius.

For a given change in electron density $\Delta n_e$, the column density gradient can be approximated as $\nabla_x N_e \simeq \Delta n_e (Z/W)$, and hence one can estimate $Z=W\nabla_xN_e/\Delta n_e$.
Scaling to the electron density typical of observed nebular filaments, $\Delta n_e \simeq 1000 \textrm{cm}^{-3}$ \citep{osterbrock57}, we infer,
\begin{equation}
  Z = 45\,W
  \left( \frac{\delta}{1\arcsec} \right)
  \left( \frac{\Delta n_e}{1000 \textrm{cm}^{-3}} \right).
\end{equation}

Thus, we find that the structures responsible for the echoes likely have sheet-like shapes, with thicknesses $W \approx 0.1{\rm\,AU}$, and sizes $L\gtrsim 4{\rm\,AU}$ and $Z \approx 5{\rm\,AU}$.

\section{Ramifications}\label{sec:rams}

We have shown that alignment of pulses before stacking allows for very sensitive observations of echoes.
In so doing, we have revealed that echoes are extremely common.

We find that the echoes are likely caused by sheet-like structures being viewed edge on, with structures elongated in only one dimension clearly excluded.
The length scales of the sheets, of $\gtrsim\!4{\rm\,AU}$ on the sky, are below the resolution of present observations of nebular filaments ($\sim\!100{\rm\,AU}$ with \emph{HST}), but given that these are thought to be the result of Rayleigh-Taylor instabilities \citep{hester08}, might well be substructures of those filaments.
Indeed, that the structures are part of a larger filament is also suggested by the clustering of the curvatures of many of the echoes, which we found requires sharing similar orientation and distance.

A possible interpretation is that the structures causing the echoes are nearly parallel ripples or smaller Rayleigh-Taylor fingers of a larger structure, perhaps a nebular filament, that happens to be crossing our line of sight.
Physically, the echoes might arise in ionized interfaces of the neutral interiors of these structures with the hot pulsar wind material.
This would naturally be sheet-like and, when viewed from the side, be able to to bend light sufficiently to cause echoes.
We would also expect these to therefore be coincident with changes in the dispersion measure, as was observed in the 1997 event reported by \cite{backer00} and \cite{lyne01}.

In this picture, one would expect that the typical curvature of the echoes would change only gradually, until, as the pulsar moves, it leaves larger filaments behind and make its way to new ones, with new typical distances and orientations, and thus different typical curvatures (see Fig.~\ref{fig:sketch_filaments}).
Indeed, from previous work, we know that the echo curvatures can vary greatly.
The 1997 echo evolved with a curvature of $2{\rm\,\mu s/day^2}$, much lower than the curvature of any echo that we see (and hence lasted much longer, though it was also far more prominent).
This is consistent with the Crab now having moved well past any structure associated to the 1997 event.

Over the time frame observed in this data, we see the 2012 December echo previously identified in \cite{driessen19}, but not that seen in 2013 January.
For the 2012 December event, our observed delay at $610\text{MHz}$ appear shorter, $\sim\!700{\rm\,\mu s}$, than that seen at $350{\rm\, MHz}$, $\sim\!1350{\rm\,\mu s}$.
\cite{backer00} showed such differences in delay may be expected for both refraction and dispersion-based scenarios; better frequency coverage would be needed to distinguish them.
It is not clear why we do not observe the 2013 January echo seen by \cite{driessen19}; perhaps at our higher frequency it could no longer bend light sufficiently.
Conversely, that \cite{driessen19} did not see many of our smaller echoes is not surprising: they looked at folded pulse profiles rather than stacked, aligned giant pulses and hence would not be able to see our typically fainter echoes at short delays.

Indeed, we have not fully used the improvement in sensitivity to echoes that we get from stacking of giant pulses: in this paper, we focused on only the visually most obvious echoes, but it is clear that there are many weaker echoes as well.
To study these would require a consistent approach to identifying them, e.g., using a Hough transform for parabolae with a range of curvatures.
Likely more significant further improvements would come by observing with larger bandwidth.
This would allow one to analyze the frequency dependency of echoes directly, and also increase the sensitivity to fainter events, thus allowing stacks of shorter duration and reduced smearing of echo profiles.
We plan to pursue this with daily monitoring of the Crab with the Canadian Hydrogen Intensity Mapping Experiment \citep[CHIME;][]{CHIME}.

\textit{Acknowledgements}: We thank the Toronto scintillometry group for useful discussions.
M.Hv.K. is supported by the Natural Sciences and Engineering Research Council of Canada (NSERC) via discovery and accelerator grants, and by a Killam Fellowship.
Pulsar research at Jodrell Bank Observatory and the use of the 42ft telescope is supported by a consolidated grant from the UK Science and Technology Facilities Council (STFC).
Computations were performed on the Sunnyvale computer at the Canadian Institute for Theoretical Astrophysics (CITA).

\textit{Software}: Software: astropy \citep{astropy13, astropy18, astropy22}, numpy \citep{numpy20}, matplotlib \citep{matplotlib07}, scipy \citep{scipy20}, tempo2 \citep{tempo12}.

\bibliography{main}{}
\bibliographystyle{aasjournal}

\end{document}